# How Do Developers Deal with Security Issue Reports on GitHub?

Noah Bühlmann
Software Composition Group, University of Bern
Switzerland
noah.buehlmann@students.unibe.ch

Mohammad Ghafari
School of Computer Science, University of Auckland
New Zealand
m.ghafari@auckland.ac.nz

## ABSTRACT

Security issue reports are the primary means of informing development teams of security risks in projects, but little is known about current practices. We aim to understand the characteristics of these reports in open-source projects and uncover opportunities to improve developer practices. We analysed 3 493 security issue reports in 182 different projects on GitHub and manually studied 333 reports, and their discussions and pull requests. We found that, the number of security issue reports has increased over time, they are resolved faster, and they are reported in earlier development stages compared to past years. Nevertheless, a tiny group of developers are involved frequently, security issues progress slowly, and a great number of them has been pending for a long time. We realized that only a small subset of security issue reports include reproducibility data, a potential fix is rarely suggested, and there is no hint regarding how a reporter spotted an issue. We noted that the resolution time of an issue is significantly shorter when the first reaction to a security report is fast and when a reference to a known vulnerability exists.

## 1 INTRODUCTION

Open Source Software development has become impressively popular in recent years. For instance, GitHub, the leading software development platform worldwide, has more than 40 million developers who have closed 20+ million issues only in 2019 [6]. The advancements in Open Source Software have encouraged the software industry and large companies such as Google and Facebook to open source their otherwise proprietary software for reasons such as engaging in a large community, obtaining prompt responses, and getting fast feedback [9].

However, security issues are prevalent in Open Source Software [4, 5, 7]. HeartBleed [2] and the Equifax [10] Breach are two remarkable examples of vulnerabilities in such projects. The former vulnerability, located in the OpenSSL library (CVE-2014-0160), exposed an enormous number of secrets to the Internet, and the latter, located in the Apache Struts 2 (CVE-2017-5638), leaked the private records of more than 140 million customers of Equifax.

Timely reaction to security issues has received great attention, but previous work has mostly revolved around discerning security issues from non-security bugs [8], or predicting the abundance of vulnerabilities in an application [15]. In the end of the day, it is the developer's responsibility to discuss and resolve security issues. Nevertheless, there is dearth of evidence regarding how developers deal with security issue reports in practice. Therefore, we investigated *the prevalence of security issue reports, their characteristics, and how they progress*. We believe this is an important step to understand the dynamic of development teams, and to uncover improvement opportunities to form a better security culture in our community.

We collected a dataset of nearly 250 000 issue reports from 182 different GitHub Java projects and analysed 3 493 security issue reports, and we inspected a significant subset of these reports. We explored the prevalence of issues, reaction time, reporter, and resolution time, among others; and when needed, we compared them to non-security issues to put our observations in perspective. We publicly shared our full dataset to facilitate future investigations [1].

We found that the number of security issue reports has increased over time, developers have succeeded to shorten the resolution time of these issues, but a great number of security issues are pending for a long time. We noted that the report of the first security issue has shifted to earlier development stages, nonetheless, security issues are still reported significantly later than non-security issues in a project. We identified that, compared to non-security issues, only a small group of developers are involved in reporting and resolving security issues. We found that a small subset of security issue reports includes reproducibility data, a potential fix is rarely suggested, and there is no hint regarding how a reporter spotted an issue. Finally, we noted that the resolution time of an issue is significantly shorter when the first reaction to a security report is fast and when a reference to a known vulnerability exists.

The remainder of this paper is organized as follows. In section 2, we present the basics of issue tracking on GitHub. We explain our methodology in section 3. We present our results in section 4 and discuss them in section 5. We report threats to validity of this work in section 6. In section 7, we give an overview of related work, and we draw our conclusion in section 8.

## 2 BACKGROUND

Issue reports are the bug tracking functionality of GitHub. They have their own section in every repository that has issue tracking enabled. Generally, any user who has read access to a repository can submit an issue. For public open source projects, this means that every GitHub user can create an issue. In order to do so, one has to fill out a simple form that contains the title and description fields. For the description, a project may provide issue templates that have a predefined structure (*e.g.*, a section for reproducibility information) that the issue reporter should follow because it is desired by the community.

Participants in an issue can have one of the following associations with the project:

- The *Owner/Member* is the GitHub user who created the repository. If the repository was created and is maintained by an organization, then all developers who belong to that organization fall into the member category.
- *Collaborators* are users who were invited by the project owners to contribute to a project and have write access to the repository.
- *Contributors* are users who do not have collaborator access to a repository, but they have contributed to a project through making a pull request that was merged into the repository.
- Users who do not belong to one of the above categories fall into the *None* category.

*Labels* are a way to organize issue reports in a project. They are predefined categories defined per project that can be applied to issues and pull requests. Each issue can have one or multiple labels. Labelling and assigning of issues only happen after the creation of an issue and can only be performed by GitHub users with the status owner, member or collaborator for the project, *i.e.*, developers with write access to the repository. Finally, an *assignee* is a person who is responsible for the issue. One or multiple users can be assigned to an issue. They receive a notification in GitHub when they are assigned to a new issue.

## 3 METHODOLOGY

We followed an empirical approach to shed light on how developers deal with security issues reports in open-source projects. In the rest of this paper, we use the terms "issue" and "issue report" interchangeably, which both refer to issues that are reported in GitHub projects. In terms of the "significance" of our observations, we rely on a non-parametric Mann–Whitney U test to judge the difference of means, and we determine the significance of a correlation based on the p-value of the Pearson's correlation.

**Selection of projects.** We had to make a selection of GitHub projects (repositories) as, practically, we cannot analyse all the projects. Therefore, we set up several selection criteria on these projects. Precisely, we stuck to Java projects as it is one of the top programming languages, it has a very large user base on GitHub, and we are very familiar with this programming language. We did not include forks of another repository to prevent duplicate projects and redundant issue reports. We selected projects that had a size of more than 2 kB to exclude repositories that were mostly empty. We did not include toy and personal projects that are not known to other developers. Specifically, we considered projects that had more than 10 forks and 10 stars, and at least 50 commits and 10 issue reports. We only selected projects that had at least one push within the last 365 days. We added this criterion to exclude inactive projects. Finally, we selected projects that used English as their issue tracking language to be able to carry out a manual analysis of the issue reports.

With these criteria, out of all GitHub projects, we were left with 5 572 repositories that we further processed to exclude those that did not include any issues of security-related type. We used a simple label-based approach to classify issue reports into security and non-security issues. For each remaining repository, we extracted any labels that contained the string "security". We found 276 repositories that had at least one such label available, and in 182 instances such a label was actually used to tag at least a single issue report in the project. We therefore proceeded with those 182 projects because we only wanted to include projects in our analysis that had both security and non-security issues.

**Full dataset.** We downloaded all the issue reports in these projects and classified those reports that were labelled with one of the aforementioned security labels as security issues and all other issues as non-security issues. We ended up with our final dataset of 182 GitHub projects containing a total of 249 043 issue reports, of which 3 493 were labelled as security issues, and a total of 852 341 discussion comments. Our data represents a snapshot at the 11.04.2020 14:00 UTC. The key statistics of these projects are summarized in table 1.

We relied on 39 features to acquire a clear insight into our full dataset. In particular, 15 features were about the repositories; 20 features concerned the issue reports; and four features were related to the comments. The full dataset as well as the explanation of each feature is available online [1].

**Sampling.** Manual analysis of all the security issues in our dataset would have exceeded our time and resources. Therefore, we drew a significant sample from the full dataset and performed our manual investigation on that sample. We were particularly interested in the discussions of the security issues, and therefore we excluded all issues with zero comments, where a meaningful analysis of the discussion was not possible. With such restrictions, our population of 3 493 security issues shrank to 2 354 commented security issues. We wanted to achieve a confidence interval of ±5% at a confidence level of 95%, which resulted in a minimum required sample size of 331.

The distribution of our population with respect to the number of comments and the status of an issue vary (see table 2). Therefore, we adopted a stratified-random sampling to preserve the distribution of the aforementioned features in our sample set and maintain sufficient representativeness to draw conclusions for the main population. We distributed the samples to the strata according to each quartile of data and ended up using a sample of 333 security issues in order to accommodate for rounding imprecisions. The composition of this sample set is shown in table 3. Finally, we used a true random number generator[1] to randomly select the number of issues from our different strata of the population.

---

[1]https://www.random.org

| Metric | commits | forks | stars | issues | pull requests | age [years] |
|---|---|---|---|---|---|---|
| Mean (Std. Error) | 6 135 (679) | 629 (123) | 1 751 (335) | 1 368 (178) | 1 416 (222) | 5.20 (0.18) |
| Std. Deviation | 9 167 | 1 657 | 4 524 | 2 397 | 2 999 | 2.42 |

Table 1: General statistics of the 182 repositories in the dataset

| Quartile | Accepted | Pending | Rejected | Total |
|---|---|---|---|---|
| 1 | 321 (13.6%) | 162 (6.9%) | 267 (11.3%) | 750 |
| 2 | 357 (15.2%) | 120 (5.1%) | 138 (5.9%) | 615 |
| 3 | 165 (7.0%) | 142 (6.0%) | 122 (5.2%) | 429 |
| 4 | 247 (10.5%) | 139 (5.9%) | 174 (7.4%) | 560 |
| **Total** | 1 090 | 563 | 701 | 2 354 |

Table 2: The distribution of 2 354 commented security reports in each quartile (number of comments)

| Strata | Accepted | Pending | Rejected | Total |
|---|---|---|---|---|
| 1 | 45 | 23 | 38 | 106 |
| 2 | 50 | 17 | 19 | 86 |
| 3 | 23 | 20 | 17 | 60 |
| 4 | 36 | 20 | 25 | 81 |
| **Total** | 154 | 80 | 99 | 333 |

Table 3: The distribution of 333 sampled issue reports based on stratification

We inspected the 333 randomly selected issue reports, the 1 335 comments in the issue discussions and the 124 pull request discussions related to these issues.

**Categorization of resources.** We categorized resources that were provided either directly in the issue or via a hyperlink into the categories "fix", "documentation", "cve" and "reproducibility". A "fix" is generally a concrete suggested change to the source code that is likely to resolve the issue. "Documentation" refers to further information about the functionality of software or a third-party library. "CVE"[2] includes all kinds of entries in public vulnerability databases. We used the value "reproducibility" if the report included any kind of resources that describe how to reproduce an issue. Finally, we assigned the value "none" if none of the information mentioned before was provided. We also checked whether the above information exists in comments.

**Inspection of pull requests.** We checked if a pull request was directly linked to an issue. If we could not find any pull request, we assigned the value "no". If a pull request existed, we checked whether or not (and how many times) the review feature on GitHub was used on the pull request corresponding to the issue. If no review had been performed, we assigned the value "no review". If one review had been conducted, we assigned the value "single review", and if more than one review had been done, we assigned the value "multiple reviews".

We checked whether the author of an issue also created the pull request, was assigned as a reviewer of the pull request, or was not involved at all in the pull request. These roles are exclusive, as the creator of a pull request should not review his or her own pull request.

We counted distinct people who left a comment on the pull request. This should not be confused with the number of participants in a pull request that is displayed on the GitHub. In fact, the former only includes participants who contributed to a pull request discussion, whereas the latter count also includes people who merely performed an action on a pull request (e.g. adding a label or merging the pull request).

We were interested to figure out how challenging it is for developers to discuss a pull request. We took into consideration how long it took to reach (i) the final merge state, or (ii) the decision to merge or abandon the pull request; and checked how strong the participants disagreed or argued. We assigned a value "low" if there was basically no disagreement between the participants of the discussion, "medium" if there was a moderate discussion going on and "high" if it was extremely challenging and controversial to come to a conclusion.

**Pilot study.** We conducted a pilot study to mitigate subjective influence during our manual study *i.e.,* to ensure what we extract during the manual inspection is correct. We randomly selected 17 security issues from the remaining 2 021 commented security issues that were not used for the actual study. The two authors of this paper inspected every issue, the comments, and its pull requests. They compared their results (*e.g.,* the types of information contained in each report) in the end and discussed them until consensus was reached.

## 4 RESULTS

We report various characteristics of security issue reports that we identified, and when needed, we compare them to non-security issue reports to put our observations in perspective.

***How prevalent are issues, and when do they emerge?*** Our full dataset consisted of 182 GitHub projects which included a total of 249 043 issues, of which 3 493 (1.40%) are security issues and 245 550 (98.60%) are non-security issues. When analysing the percentages of security issues in each individual project, we realized that security issues generally make up a very small percentage of all the issues of a project. We found that 75% of all projects have a percentage that is lower than 2.23%.

> Security issues, expectedly, comprise only a small proportion of all issues in a project

We were interested to know when security issues emerge in a project. We analysed the time interval between the creation of a project and the creation of the first security issue, and also the mean time interval between subsequent security issues in a project. We found that the median is 655.50 days (interquartile range (IQR) 1 202.25 days) until the creation of the first security issue. Once the first security issue is created, we found that the median time between subsequent security issues is 54.37 days (IQR 107.06 days). To put these results into perspective, we also calculated the numbers for non-security issues and found that the median is 28.21 days (IQR 140.03 days) until the creation of the first non-security issue. Once the first non-security issue is created, we found that the median time between subsequent non-security issues is 2.53 days (IQR 4.68 days). This shows that security issues are created later in a project and emerge less frequently than non-security issues.

---
[2]Common Vulnerabilities and Exposures

> *Security issues are reported much later than non-security issues in a project.*

We were wondering if the activity of a project is related to the number of security issues it contained. We used three different metrics to assess the activity. The number of commits is an indicator for the activity in terms of code changes, and the number of forks and the number of stars give an idea of how many developers use a project. We applied Pearson correlation coefficient analysis to understand the relationship between the number of security issues and these metrics. We found that there is a statistically significant positive correlation between the number of security issues and all three metrics, with the 2-tailed p-value being smaller than 0.001. According to Cohen [3], the correlation coefficient indicates that the effect is medium for the number of commits (r=0.374) and strong for the number of forks (r=0.552) and stars (r=0.593). We can say that the more active a project is, the more security issues it contains or vice versa.

> *The more active a project is, the more security issues it contains, or vice versa.*

**How often do developers report issues, and what is their relationship with the projects?** We counted how many security issues and non-security issues reporters posted in the analysed projects. We found that for security issues, the mean is 2.37 (SE 0.40) reported issues per reporter, and for non-security issues, the mean is 2.61 (SE 0.72), showing that there is no significant difference. We analysed how far reporters of issues in general and security issues in particular overlap. We found 30 490 distinct reporters of issues, of which 29 749 (97.6%) reported only non-security issues; 130 (0.4%) reported only security issues; and 611 developers (2.0%) reported both security and non-security issues. Therefore, only 741 of 30 490 developers (2.4%) report security issues. Additionally, we found 15 cross-project security issue reporters, *i.e.*, they created security issues in more than one of the 182 projects in our dataset. Thirteen of those reporters reported security issues in two projects, while two reported in five projects.

> *Only 2.4% of issue reporters report security issues. Most of these developers also report non-security issues.*

We analysed how the reporters of security and non-security issues are associated with the project the issue belongs to and found significant differences. For security issues only 733 of 3 493 (21.0%) issues are reported by outside users who are not associated with the project at all, 1 257 of 3 493 (37.5%) come from core members (members and collaborators) and 1 448 of 3 493 (41.5%) from established contributors to the project. For non-security issues a lot more issues are reported by unassociated users, namely 86 325 of 245 550 (35.2%). Here, core members account for 70 626 of 245 550 (31.1%) and contributors for 82 684 of 245 550 (33.7%) of non-security issues.

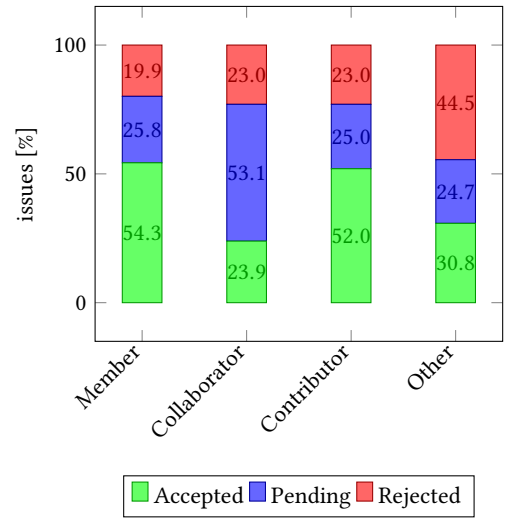

Figure 1: The performance of issue reporter groups

Comparison between accepted and rejected security issues revealed that accepted issues are more often reported by core members of a project than rejected issues; and rejected issues are reported much more often by unassociated users than accepted issues. Digging into the status of issues reported by each user group, shown in figure 1, proves that unassociated users have the worst performance *i.e.,* the proportion of issues that they reported and were rejected is two times more than for other issue reporters. Members and contributors have the best performance as more than half of their reports were accepted. Unexpectedly, more than half of issues that collaborators reported were pending, which is twice more than for other groups.

> *The role of established contributors is significantly more in the reporting of security issues than non-security issues, whereas it is the opposite for outsiders.*

**How long are issue discussions, and how many developers participate in such discussions?** We counted how many comments exist in issue discussions, and we found that security issue discussions have a mean of 2.87 (SE 0.09) comments, which is less than non-security issue discussions that have a mean of 3.47 (SE 0.01) comments. Surprisingly, 736 out of 2 527 concluded security issues (*i.e.,* 29.1%) did not receive any comment. When analysing the data per project, we found that in 81 out of 182 analysed projects (44.5%) security issue discussions contained on average fewer comments than non-security issue discussions in the same project. We did not note any significant difference in the number of comments on accepted security issues v.s. rejected security issues.

> *The majority of security issues contain a maximum of three comments. Unexpectedly, 29% of security issues are closed without any discussion.*

We analysed how many developers participate in discussions of issues. We excluded pending issues and only included closed issues into our analysis because a lot of very recently created issues do not yet have any participants in their discussions. We did not find any significant difference between the number of distinct participants in security issue discussions and non-security issue discussions (1.51 vs 1.55).

We also studied how many developers participate in the security issue discussions of a whole project. We found that on average only 7.78 (SE 1.47) developers are involved in all security issues of a project combined. The median is even lower with only three developers. This is very few, especially when we compare it to the number of participants in non-security discussions of a whole project, where we find that on average 231.61 (SE 42.61) developers are involved in all issues of a project combined, with the median being 63 participants. We were wondering how far participants in issue discussions in general and in security issue discussions in particular overlap. In our dataset we found 37 093 distinct participants of issue discussions. Only 941 of 37 093 (2.5%) developers participated in both security and non-security issue discussions. 35 904 (96.8%) participated only in non-security issue discussions and 248 (0.7%) only in security issue discussions. This leads us to the conclusion that only 1 189 of 37 093 developers (3.2%) are involved in security issue discussions. Furthermore, we checked how frequently developers participate in issue discussions. For each participant in security issue discussions, we calculated the ratio between the total number of security issues in a project and the number of security issues in a project that he or she participated in. We found that the developers participate in more than one-fourth of the security issue discussions of a project with the mean ratio being 0.25 (SE 0.01). Of course, this value is extremely lower for non-security issues, where the mean participation ratio is only 0.0083 (SE 0.00029).

> *We learned that of total participants in issue discussions, 3.2% are involved in security discussions. In each project, on average, 7.78 (SE 1.47) developers participated in all the security issue discussions, which is greatly lower than participants in non-security issue discussions. However, participants in security issue discussions take part in such discussions much more frequently than participants in non-security discussions.*

**How fast do developers react to issues and resolve them?** We wanted to know how fast developers react to security issues. We analysed two metrics to answer this question. We looked at the time it takes until the first comment emerges on an issue report, often called "reaction time", and computed the mean time it takes between every two consecutive comments that follows after the first comment in the discussion. Naturally, issues without any comments were excluded from this analysis.

Judging by the mean, non-security issues get their first comment approximately six days faster than security issues (47.52 days vs. 53.17 days). Standard deviation is very high with this metric indicating that the data is very spread out and outliers, *e.g.*, issues taking years until their first comment, are influencing the mean. However, the median reveals a different picture. 50% of security issues receive the first comment nearly one hour earlier than non-security issues. The comparison of mean time between subsequent comments in the issue discussion also suggests that the discussions proceed slower in security issues than in non-security issues (median 6.87 days vs. 4.71 days).

We also compared these two metrics between accepted, pending and rejected security issues. We found that rejected security issues have a much slower reaction time than accepted security issues, which is less than half the time with 35.89 (SE 4.48) days vs. 76.80 (SE 7.77) days. Also, the mean time between discussion comments is significantly longer in pending and rejected security issues than it is in accepted security issues.

> *The first reaction to an issue is within a few hours. Nevertheless, discussions related to security issues progress slower than that of non-security issues. In security discussions, the first reaction to an issue as well as the follow up discussions happen much faster in accepted issues than in rejected ones.*

We looked into the resolution time of security issues. We observed that the data is very spread out with a standard deviation of 283.51 days. 25% of the security issues are resolved within 1.90 days or less, and 25% take 94.88 days (approximately three months) or more to be resolved. When judged by the mean, security issues are resolved slightly faster than non-security issues (110.08 days vs. 116.97 days), when judging by the median; however, non-security issues are resolved in half the time of security-issues (8.06 days vs. 15.32 days). When analysing the data per project, we found that in 94 out of 182 available projects (51.6%) security issues were on average resolved slower than non-security issues.

We found a strong correlation (r= 0.637, p= 0.000, n= 1791) between the resolution time of a security issue and its reaction time, *i.e.*, the time until the first comment. This signifies the importance of the first comment in a security issue. The faster there is a first comment, the faster a security issue is resolved or vice versa. We noticed that accepted security issues are resolved significantly faster than rejected security issues with a mean of 84.95 (SE 5.03) vs. 153.32 (SE 9.61) days. Note that the mean age of pending security issues is 625.94 (SE 17.84) days.

> *Developers are mostly slower in concluding a security issue report than a non-security issue. In at least half the cases, the average resolution time of a security issue is twice longer than a non-security issue. Reports with shorter reaction time are concluded faster. Accepted security issues are concluded much faster than rejected ones.*

We also analysed how long issues without a single comment have been pending, and how long the resolution time for accepted and rejected issues with zero comments was. We found that in such cases non-security issues have been pending only insignificantly longer than security issues with means of 614.65 (SE 5.18) days vs. 602.29 (SE 28.28) days. For closed issues without any comments the resolution times were on average 34.58 (SE 0.67) days and 33.49 (SE 4.17) days for accepted, and 132.02 (SE 3.50) days and 48.27 (SE 11.92) days for rejected non-security and security issues respectively. This shows that issues without any comment get rejected significantly faster in the case of security issues than for non-security issues.

We also studied the resolution time of security issues that were concluded without any comment, and interestingly, found that such issues are resolved significantly faster than security issues that received at least one comment. In particular, accepted issues without comments were resolved in 33.49 (SE 4.17) days, whereas accepted issues with comments took 108.94 (SE 7.00) days to conclude. In the same vein, rejected issues without any comment were resolved in 48.27 (SE 11.92) days, whereas those with comments took 187.48 (SE 11.86) days to conclude.

> *The resolution time of security issues that did not receive any comment is significantly faster than those with comments. Pending security issues that have not received any single comment have an average age of more than 600 days, which surprisingly, is too long.*

**How often do developers get assigned to issues, and how do they perform?** We found that in 1 698 of 3 493 (48.6%) security issues at least one developer was assigned to the issue, and for non-security issues this share was lower with 107 923 of 245 550 (44.0%). We find that 1 069 of 1 598 (66.9%) accepted security issues have at least one assignee, while 316 of 966 (32.7%) pending issues and 313 of 929 (33.7%) rejected issues have an assignee. We conclude that accepted issues have most assignees, while in pending and rejected issues about two-thirds of the security issues do not have an assignee. This is also confirmed by the mean values with 0.71 (SE 0.01) assignees for accepted, 0.31 (SE 0.02) for pending and 0.33 (SE 0.02) for rejected security issues. In reverse, this also yields a much higher acceptance rate for security issues that have an assignment with 77.4% vs. 46.3% for issues with no assignment. In 63.3% of security issues with assignees, actually a self-assignment took place, *i.e.,* a user assigned himself to a security issue. Finally, we found that neither the reaction time nor the resolution time differs significantly between issues with and without an assignee.

> *The acceptance rate of security issues with an assignee is much higher than issues with no assignee.*

Further analysis showed that 328 distinct developers were assigned to the 1 698 security issues that had at least one assignee. We found that these 328 developers (in short, "security assignees") reported 1 505 of 3 493 (43.1%) of the security issues in our dataset. The mean number of reported security issues by such a security assignee is 7.00 (SE 0.80), which is significantly higher than the mean number of 2.37 (SE 0.40) security issue reports by security issue reporters in general. Further investigations revealed that of the 1 117 developers who were not security assignees, 1 079 (96.6%) have fewer comments in security issues and fewer security issue reports than the average security assignee. We noted that the mean resolution time, 80.86 (SE 5.95) days, is significantly shorter for security issues reported by security assignees than the global resolution time for security issues, which is 110.08 (SE 4.80) days. Furthermore, we also found that the acceptance rate for security issues reported by security assignees is 77.7%, which is much higher than the global acceptance rate for security issues of 63.2% and the acceptance rate of non-security assignees which is 49.0%.

Finally, we also analysed when the first assignment of a security issue happens. On average, the first assignment takes place 51.00 (SE 4.52) days after the creation of a security issue. It happens earlier for accepted issues with a mean of 39.29 (SE 5.30) days than for pending and rejected issues with means of 87.19 (SE 13.10) and 54.49 (SE 9.69) days respectively. At the time of the first assignment, a mean of 1.34 (SE 0.09) comments have already been left on the security issues, and a mean of 2.81 (SE 0.11) comments follow after the first assignment, which shows that, on average, the amount of discussion would significantly increase after an assignment.

> *Security assignees, on average, report significantly higher number of security issues and have a higher acceptance rate than security issue reporters in general. Besides that, the average resolution time of security issues that are reported by such developers is significantly shorter than the global resolution time of security issues.*

**Have characteristics of security issues changed in the past years?** We examined the number of security issues over the course of the last six years from 2014 until April 2020.[3] A big increase in the number of security issues after 2016 was evident. We noted that the proportion of accepted issues is consistently bigger than pending and rejected issues.

We also analysed how security issue evolve when they age. We grouped the 3 493 security issues in our dataset into ten "age bins", according to the 10%-quantiles of the age of the issues. In general, as security issues age the numbers of accepted and rejected issues decrease (because they are closed) and the number of pending issues increases of course. For issues between two and 47 days old, there are more than twice as many accepted issues than rejected issues. Notably, the number of accepted and rejected security issues converges more as the issues get older. We confirm that the cross-comparison between the age group and the issue statuses is significant. We also found a strong positive correlation (r=0.635*, p=0.000, n=182) between the age of a project and the days until the first security issue is reported. In other words, the first security issue is reported faster in "younger" projects than in older projects.

We checked the development of the mean number of participants as well as the mean number of comments in security issue discussions in the past years. We found that the number of participants who were involved in security issues has not really changed over the years. Nevertheless, in recent years fewer comments have been needed to resolve issues. Interestingly, there are constantly more comments on rejected issues than accepted and pending issues. Our further investigations show that in recent years security issues are reported more often by contributors than the core members of the projects, which could be explained with the growing open source community. Nevertheless, the participation of core members, contributors, and outsiders in the discussions has remained almost stable over the years. Additionally, we explored the distribution of the number of participants and the number of comments as security issues age and confirm that both increase as issues age. Nevertheless, if an issue is not resolved after half a year, fewer participants get interested in joining the discussion.

---
[3]We excluded issues created before 2014 as they were rare in our dataset.

We analysed if there are any trends in the resolution time of security issues over the years, and found that it has become faster and faster. Notably, rejected security issues are resolved significantly slower than accepted security issues almost all the time. We also confirm that reaction time to security issues has significantly decreased in recent years.

> *The number of reported security issues have increased over the years, but also developers have succeeded to spot, react, and resolve security issues faster. Interestingly, rejected security issues needed longer discussions and progressed slower than accepted issues. The population of issue reporters has shifted from core members to contributors.*

**What kind of information exist in security issue reports and their discussions?** We found that 200 out of 333 security issue reports (60.1%) include specific types of information shown in figure 2. It is most common that developers refer to documentation (40.2%), but reproducibility data is not very common (14.7%). Only in 4.5% of cases reference to vulnerability entries (CVE) exist. We realized that issue reporters rarely, *i.e.*, only 2 issues out of 333 (0.60%), include any concrete fix in their issue descriptions. We found that accepted security issues are more enriched than rejected ones. Most notably, accepted security issues more often refer to documentation (44.2%) or reproducibility information (16.9%) than rejected security issues (34.3% resp. 14.1%). Furthermore, there was no occurrence of a pending security issue containing a reference to a CVE entry. This cross-comparison between further information and issue statuses is significant, with a p-value of 0.027.

We also checked whether any of the aforementioned types of information exist in security discussions, and found that in 77 of 333 security issues (23.13%) a fix was provided in the comments, and in 90 of 333 issues (38.63%) comments included documentation.

We did not find any evidence that the status of an issue, its resolution time, or even the length of the issue discussion would change if the security issue report or its discussion includes a fix. Surprisingly, we neither found any evidence that a reminder (*i.e.*, a comment that asks developers if there is any update on an issue or mentions certain developers to remind them of an issue) would speed up the resolution process of a security issue.

> *The presence of a possible fix is very rare (i.e., 0.7%) in the security issue reports, and only a small subset of issues (i.e., 14.7%) include reproducibility data. We confirm that a possible fix arose in 23.13% of the issue discussions.*

**How do developers apply a security fix?** Our manual analysis revealed that 106 out of 154 analysed accepted security issues (68.8%) were fixed by a pull request, 44 issues (28.6%) were fixed with a direct commit to the repository without a pull request, and in four cases it was unknown where or how a fix was implemented. A key functionality of pull requests on GitHub is the possibility to add users as reviewers so that they can assess the proposed code changes. Fortunately, we found that 105 of 124 (84.7%) analysed pull requests were reviewed at least once by another person than the pull request creator. In 54 of 124 (43.6%) pull requests there were even multiple reviews carried out. We noticed in most cases where there is more than one review, reviews from multiple reviewers were requested simultaneously right from the beginning, because the creator wanted to have feedback from more than one person, or because the policies of a project require a certain number of reviews.

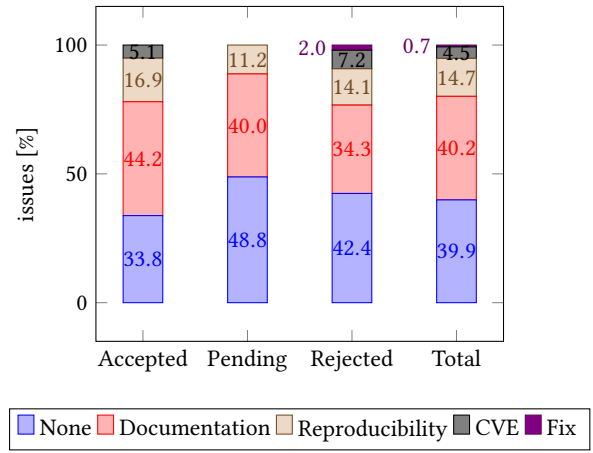

Figure 2: Further information in security issue reports

We assume that an issue reporter is knowledgeable about the issue, and therefore we investigated whether the reporter has any role in a pull request that emerged from the reported issue. We found that a security issue reporter is not involved in 50.8% of pull requests. In 35.5% of pull requests the issue reporter has made the pull request too, and in 13.7% of cases the issue reporter has reviewed the suggested fix.

> *Security fixes are often (68.8%) applied to a repository through a pull request. Most of pull requests (84.7%) are reviewed at least by one other developer. However, an issue reporter who is presumably a very knowledgeable person about the issue, is not involved in 50.8% of pull requests at all.*

**How long are pull request discussions, and how many developers participate in such discussions?** We found that the number of participants in pull request discussions that emerged from security issues follows a normal distribution with a mean of 2.09 (SE 0.11) and a standard deviation of 1.20, and only four pull requests have more than four participants. Therefore, the mean number of participants in the pull requests is significantly higher than that of the issue discussions (mean 1.51 and SE 0.03).

We also investigated the number of comments in the pull request discussions, and found that they have a mean of 10.06 comments (SE 1.48) with a standard deviation of 16.52, which shows the data is widely spread out. Hence, pull request discussions have a significantly higher number of comments than the issue discussions (mean 2.87 and SE 0.09). Moreover, we found that the number of comments in the issue discussion has a statistically significant correlation of medium effect with the number of comments in the pull request discussion (r= 0.427, p= 0.000, n= 124). In other words, the more comments exist in the issue discussion, the more comments there are in the pull request discussion or vice versa.

The main pattern that we noted about the issues that had very extensive pull request discussion is the following. A pull request is opened because someone has "coded" a possible solution for the security issue. This solution however is either incomplete, does not pass automated testing, has merge conflicts or there are simply changes that are requested/proposed by a fellow developer before it is merged. This leads to changes in the pull request and discussion about those changes. Manual investigation has also shown that this effect is amplified if the pull request is very complex or challenging, which often causes "back and forth" discussion between developers.

> *The number of participants as well as the number of comments are significantly higher in pull requests than issue report discussions.*

In cases where pull requests had zero comments, we noticed that the changes were apparently quite straightforward or minor and there was no need for discussion. Those cases did not have many comments in the issue discussion either. We noticed though, even when there is no discussion, the GitHub review feature was used in about half of cases, showing that there was some form of feedback.

*What factors possibly influence the resolution time of security issues?* Figure 3 presents cases where a strong correlation exists between the mean resolution time of issues and several features that we extracted in our manual analysis. We found that resolution time is significantly longer if documentation is provided in the comments; the issue is reported by core members (members and collaborators) or vice versa. Conversely, security issues conclude significantly faster if a CVE reference exists in the report or the comments; the pull request is not reviewed multiple times or vice versa. As we only measured correlation with our study, future work is necessary to identify if and how much each of the above factors actually impact the resolution time.

In our quantitative analysis we found that the faster developers react to an issue, *i.e.,* leave the first comment, the faster a security issue is resolved or vice versa. Nevertheless, we were not able to identify any strong indicator to discern security issues that attract a quick reaction from those that have a slow reaction time. We believe an interview study with developers is necessary to better understand these phenomena.

> *The resolution time of a security issue is significantly shorter if a CVE reference exists, or the issue reporter is involved.*

## 5 DISCUSSION

We studied the distribution of security issues that were reported from 2014 until 2020, and found a big increase in the number of reported issues in recent years, particularly starting in 2017. We noted that, once the first security issue of a project is created, a new one appears on average every 54 days. Luckily, security issues are reported much earlier in more recent projects than in older projects. These observations might indicate the integration of security in development workflows in newer projects, which is different from older projects that apparently treated security as a single goal (*i.e.,* check security at a certain point in development). Moreover, the rise in the adoption of open-source software by organizations as well as the growth of enterprise open source (*i.e.,* open-source software that is backed by enterprises) might have improved attention to security concerns over the last few years.

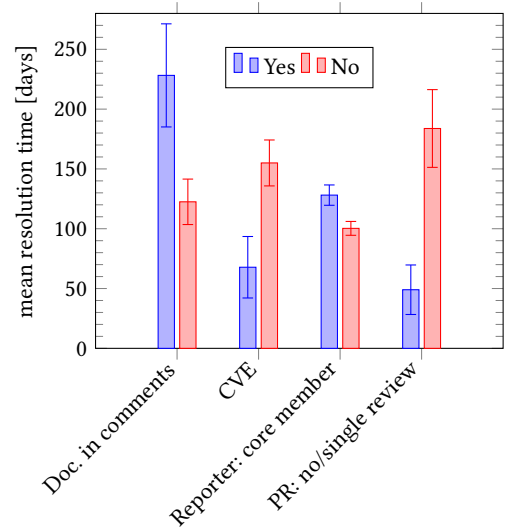

Figure 3: Mean resolution time of closed security issues for different values of categorical features

Security issues are reported by a very small circle (*i.e.,* 2.4%) of developers. We believe security issues, as opposed to non-security issues, are not observable to many developers and spotting such issues requires domain expertise that only a limited number of developers possess. This is in line with our observation that security issues are mostly reported by developers associated with the project (*i.e.,* core members and established contributors), as opposed to non-security issues, which are greatly reported by those who are not associated with the project. We identified a strong correlation between the popularity of a project, in terms of the number of forks and stars, and the number of security issues in a project. This finding would lead to the hypothesis that the more developers are involved in a project, there will be more opportunities to discover security issues or vice versa. Likewise, a security issue identified in a fork project may get reported back to the main project. Surprisingly though, we found that half of security issues reported by contributors are pending. Future work may investigate which issues tend not to get resolved. Are they mostly low-risk issues, or are they high-risk issues that remain open?

We identified a visible decrease in the number of comments and the resolution time of security issues over time. We speculate the movements toward agile methodologies may have resulted in resolving issues more efficiently (*i.e.,* shorter time and discussion). One other reason might be the availability of more tool support. For instance, seven cases used the GitHub Security Advisory that enables periodic scanning of the source code to detect known vulnerability patterns in the project and its dependencies. Otherwise, we noticed that developers do not mention any specific tool that might have helped them to identify the issues. Remarkably, we observed that many security issues are concluded without any discussion. Future research may investigate the characteristics of such issues to identify what they have differently than other issues, and encourage practitioners to do so.

We found that, on average, the same number of participants are involved in security and non-security discussions, but security discussions progress slower, and resolving them takes significantly longer than non-security issues. A possible explanation might be the complexity of security issues, which requires expertise to comprehend and address. The huge difference between non-security and security issues in terms of the number of issue reporters, and their association with the projects confirms the scarcity of security knowledge among developers. Besides that, security issues are not as straightforward as non-security issues to spot and test, which further underlies the slow progress and resolution of security issues. We found that developers who are assigned to security issues play an important role in resolving security issues. They outperform 97% of the other developers that are involved in security issues by reporting more issues and participating in more discussions. These security assignees were involved in roughly 64% of all security issues, and 77.7% of the issues that they reported were accepted, which is much higher than the other security issue reporters. Future work may investigate how we can boost security discussions. Particularly, under what circumstances developers participate in addressing security issues and how we can streamline the participation of more developers in this process.

We noted that in security issues, the longer, in terms of the number of comments, an issue discussion is, the longer a pull-request discussion will be or vice versa. This is consistent with the intuition that a hard or complex issue that requires more discussion also leads to more discussion during the implementation phase (pull request) because the changes are likely more complex as well. What's more, pull-request discussions are significantly more involving, in terms of the numbers of comments and distinct participants, than issue discussions. A potential explanation derived from our observations is that an issue discussion is often about examining the validity of the report and providing the outlook of a possible fix, whereas a pull request discussion is mostly about a concrete solution and its impact on the program, requiring much more "going back and forth" discussions if the matter is complex or hard to fix.

We analysed the relationship between the resolution time and various characteristics of security issues, and discovered a number of significant relationships. Firstly, the shorter the initial reaction time to an issue is, the faster developers resolve that issue or vice versa. This finding could suggest, not prove however, that reacting to an issue quickly may generate enough attention such that the issue is also resolved promptly. However, future work must investigate under what circumstances developers reaction to a security report would increase. For example, is it dependent on the way an issue is explained or is it due to a perceived level of severity by developers? We also discovered that the resolution time is shorter if the issue reporter is a core member of the project or vice versa. Recognition and trust to internal members of a project are possible reasons why security issues reported by them are resolved faster. Moreover, we found that the resolution time for issues that were reported by "security assignees" was significantly shorter than the global mean. These developers were involved in a large proportion of security issues, which reflects their experience and commitment to resolving security issues.

The resolution time of security issues is also shorter if a CVE report is referenced in the issue report/discussion or vice versa. This indicates that a CVE is alerting developers and is a sign of urgency of a security issue, however only 4.5% of the security issues included a reference to a CVE. Surprisingly, we discovered that the resolution time of security issues is longer if further documentation is provided in the discussions or vice versa. We realized, during the manual analysis, that mostly complex issues require/include documentation to be understood correctly. Unexpectedly though, the issue reporter of a security issue is not involved at all in 50% of the issue's pull request, neither as creator nor as a reviewer or discussion participant of the pull request. This might be explained by the fact that those issues were straightforward to fix and did not require the reporter's expertise. Indeed, 72% of the pull requests corresponding to such issues were categorized as "not challenging" in our manual analysis.

Finally, we came across a few security issues that were special in some aspects. In two cases, a security vulnerability was not disclosed publicly ("responsible disclosure") in the issue report. In fact, the reporter did not explain the issue but asked for a contact email address where the details could be submitted. We observed two issues due to a vulnerable outdated dependency. Although the vulnerabilities even received a CVE identifier, the dependency was upgraded to a newer (safe) version after more than one year. Interestingly, in one security issue,[4] the reporter explicitly stated in the issue description that the issue is reported by a group of researchers from the University of Nebraska. We found six security issues that directly led to the creation of CVE entries. The first one, CVE-2019-11405, was created because the OpenAPI Tools OpenAPI Generator used http:// instead of https:// in various Gradle build files, which may cause insecurely resolved dependencies.[5] The five other security issues were related to the FasterXML jackson-databind project.[6] Two vulnerabilities would allow an attacker to perform remote code execution (RCE) attacks (CVE-2018-14718 and CVE-2018-14719), one would allow external XML entity (XXE) attacks (CVE-2018-14720), another one server-side request forgery (SSRF) attacks (CVE-2018-14721) and the final one would allow exfiltration of content (CVE-2018-11307).

These findings drive several research directions as follows:

- Which issues tend not to get resolved, and notably, why half of issues reported by contributors are pending? For example, are there severe issues that stay unresolved, or are they mostly low-impact bugs?
- Under what circumstances do developers participate in resolving security issues, and how can we streamline the participation of more developers in this process? For instance, does the inclusion of documentation and a fix help onboarding non-expert developers to the process?

---

[4]https://github.com/TEAMMATES/teammates/issues/8183
[5]https://github.com/OpenAPITools/openapi-generator
[6]https://github.com/FasterXML/jackson-databind

- Why are the developers' reaction to certain reports fast? For instance, is it because of the way an issue is explained, or is it due to a perceived level of severity by developers?
- Why are many issues resolved without any discussion? For instance, how are such issue reports different from slow-progressing issues, and how can we encourage the reporters to do so?

## 6 THREATS TO VALIDITY

There exists a threat to validity related to the bias in sampling projects. Especially, the restrictions to Java projects and projects with English issue tracking, which have been imposed for logistical reasons, might bias the results. The data for our study represents a snapshot of a very specific point in time. The projects will likely evolve in the future, and the results could look different if the study were to be carried out at a later time. There is also a threat to validity concerning the classification of issues into security issues and non-security issues. We used a simple label-based approach to do the classification. It was clear that we will not be able to capture all security issues with this approach, e.g. in projects where strict labelling is not enforced. Finally, the proportion of non-security issue reports was much larger than security reports in our dataset. Although such a class imbalance can be problematic for certain observations, we often relied on a t-test to compare security and non-security reports, which handles unequal sample sizes.

Finally, personal bias is also a threat to validity. We tried to describe our methodology, principles and criteria as transparently as possible and adhere to them in order to minimize the effects of our personal opinions or expectation of results. Furthermore, we tried to mitigate subjective influence during the manual analysis by conducting a pilot study previous to the real study, where results were discussed and agreed upon between the two experimenters.

## 7 RELATED WORK

Zahedi et al. investigated the prevalence of security issues in a random sample of open source projects and found that security issues comprise approximately 3% of all issues and the majority of them are related to identity management and cryptography [14]. Morrison et al. conducted an empirical analysis of three open source projects to study the differences between the discovery and the resolution of defects versus vulnerabilities [12]. They found that vulnerabilities are discovered later in the development cycle and are resolved more quickly than defects. Pletea et al. analysed security discussions on GitHub with the specific intent to extract emotions from the comments [13]. They found that more negative emotions are expressed in security-related discussions than in other discussions, which confirms the anecdotal evidence that implementing application security can often lead to frustration and anger among developers. Meyers et al. analysed the style and the tone of developers' language in bug reports in the Chromium project [11]. Their initial investigation of metrics such as formality, informativeness, implicature, politeness, and uncertainty showed that the difference between security and non-security conversations are small but statistically significant. Zhou et al. investigated the association between bounties and the likelihood of addressing issue reports on GitHub [16]. They found that bounty issue reports are more likely to be addressed if they are for projects in which bounties are used more frequently. The bounty value of an issue report is the most important factor that is associated with the issue-addressing likelihood in the projects in which no bounties were used before.

## 8 CONCLUSION

Reacting to security issue reports in open-source projects is crucial, however, there is a dearth of evidence regarding developer practices from reporting such issues to resolving them. We conducted an exploratory study of security issue reports in 182 GitHub projects. We described several characteristics of these reports and current practices in terms of reporters, assignment, reaction, discussion, status, and resolution time, among the others. Our findings motivate the need for several research directions to better support the open-source community in resolving security issues.